\documentclass[12pt]{iopart}

\begin{document}
\title{Towards a new quantum cognition model}
\author{Riccardo Franco}
\address{riccardofrancoq@gmail.com}
\date{\today}
% ----------------------------------------------------------------
\begin{abstract}
This article presents a new quantum-like model for cognition explicitly based on knowledge. It is shown that this model, called QKT (quantum knowledge-based theory), is able to coherently describe some experimental results that are problematic for the prior quantum-like decision models. In particular, I consider the experimental results relevant to the post-decision cognitive dissonance, the problems relevant to the question order effect and response replicability, and those relevant to the grand-reciprocity equations. A new set of postulates is proposed, which evidence the different meaning given to the projectors and to the quantum states. In the final part, I show that the use of quantum gates can help to better describe and understand the evolution of quantum-like models. 
\end{abstract}
%\pacs{
%03.67.Mn, 03.65.Wj, 42.50.Dv
%}
 \maketitle
% ----------------------------------------------------------------
%%%%%%%%%%%%%%%%%%%%%%%%%%%%%%%%%%%%%%%%%%%%%%%%%%%%%%%%%%%%%%%%%5
\section{Introduction}
Subjects' behaviour is often claimed to be irrational \cite{gilovich2002heuristics}. Quantum cognition, a quite recent research field making use of concepts and methods taken from the quantum theory and quantum probability, seems to be a particularly useful and promising framework to describe a variety of seemingly “irrational” judgment and decisionmaking findings, thus providing a new perspective for cognitive science. The main hypothesis is that human reasoning - in a wide range of situations relevant to bounded-rationality - obeys the laws of quantum rather than classical probability. 
Quantum cognition allows taking into account puzzling effects like conjunction fallacies \cite{franco2009conjunction}, disjunction fallacies, averaging effects, unpacking effects, and order effects \cite{busemeyer2011quantum}, violations of sure-thing principle in decision theory \cite{pothos2009quantum},  violations of symmetry in similarity judgements \cite{pothos2011quantum} and producing new predictions like the quantum question (QQ) equality \cite{wang2014context}.
\\
Despite its successful applications, in last years some experimental data evidenced that the theory needs a deeper understanding. In particular, the violation of the Grand-reciprocity (GR) equations \cite{boyer2016testing} and the problems relevant to question order effects and response replicability \cite{khrennikov2014quantum}. An even deeper critique arrives from \cite{pleskac2013s}, where it is noted that the same formalism has been used (for example in conjunction fallacy experiments) to describe both judgements (subjective probabilities) and choice tasks (objective probabilities). The authors thus argue that \textit{to suppose these two models are of the same class seems to be an error}.

In this article, I present a new quantum-like theory for cognition based on a new set of postulates. Even if the mathematical framework is essentially the same of the previous quantum-like model, such postulates allow for a more coherent description of experimental data in terms of quantum states, preparations, test and information architecture, thus overcoming the problems evidenced before.
In particular, the new model clearly distinguishes between subjective and objective probabilities, focusing on the concept of subject's knowledge.
  I also present other set of experimental results (the post-decision cognitive dissonance case) that are in contradiction with previous quantum decision theory, while correclty described with the new set of postulates.
  
Finally, I introduce in the model concepts and mathematical tools taken from quantum computation to quickly operate with cognitive models based on quantum composite systems. These tools suggest a procedural description of cognitive processes in terms of gates and algorithms, and a cognitive interpretation of the quantum phase.

In summary, this articles evidences that the potentialities of quantum formalism in cognitive science aren't at the moment fully explored and used, and much work remains to be done.

\section{Starting from the basics}
Quantum theoretical predictions rely on conceptual models which involve elusive microscopic objects, such as electrons, photons, etc. The latter are usually considered as real things, but they occasionally display extraordinary nonlocal properties, quite at variance with intuitive classical realism. However, the predictions of quantum theory are relevant to the probability of occurrence of stochastic macroscopic events, following specified preparation procedures, such as, for instance, the triggering of particle detectors following the collision of a specified beam with a given target. 

In a similar way, quantum cognition theory makes predictions relevant to subject's anwsers following specific preparation procedures consisting in input information. Such predictions rely on another elusive object, the subject's thougths. Nobody has never seen really a thought or an electron. 

The use in cognitive psychology of definitions taken from the quantum theory offers the opportunity of getting precise concepts already built to describe situations with uncertainty, measurements and interference. Following \cite{peres2006quantum}, we recall two important definitions in the quantum theory, the \textit{preparation} and the \textit{test}. A \textit{preparation} is an experimental procedure that specifies the quantum state, like a recipe in a good cookbook. In quantum physics, any experiment is always performed over a set of identically prepared states. The \textit{test} is the process leading to the experimental outcome. In the first part, it is like a preparation, but it also includes a final step in which information, previously unknown, is supplied to an observer. This is the new information about the quantum state that the test allows one to obtain.
Note that a preparation usually involves tests, but they are followed by a selection of specific outcomes. For example, a mass spectrometer can prepare a certain type of particle by measuring the masses of various incoming particles and selecting those with the desired properties.

In cognitive psychology we can use preparation and test in a similar way. The preparation represents the information given to subjects in the preliminar part of an experiment. The test is the final part of the experiment, where subjects provide an outcome. As a trivial example, let us suppose that in the preparation phase we say \textit{there is a car, which can be blue or red}. Thus we have defined the possible outcomes of the test. The preparation can let subjects be completely uncertain about the car's color (blue or red, we don't know), or it can add a selective information, like for example \textit{we know that the car is red}. In the test phase, we ask to the subjects the color of the car. In the first case, they will answer one of the two colors with 50\% of probability. In the other case, they will all answer \textit{red}. Even repeating the question, they will give the same answer. But if we ask to subject another question like \textit{Is that car driven by a woman?}, the answer will be \textit{yes} for some subjects and \textit{no} for others. 

In the quantum theory, test on  a given quantum system is called \textit{maximal or complete} when the number of different outcomes $N$ (let us suppose that $N$ is finite) is also the maximum number of different outcomes obtainable in a test of that system. An incomplete test is one where some outcomes are lumped together, for example, because the experimental equipment has \textit{insufficient resolution}, leading to degenerate outcomes. In the quantum cognition context, we can say to subjects that the car can be \textit{red}, \textit{blue} or \textit{yellow}, but in the test we only ask if the car is \textit{red} or \textit{not red}.
It is clear that in cognitive psychology the resolution of the test is an important problem and depends on many factors: the possible outcomes presented in the preparation, but also the cognitive resources that subjects can use.

In physics, tests allow computing event probabilities. Of course, what we obtain is the  the measured relative frequency of an event, but if we repeat the same experiment a large (but finite) number of times, we can expect a small difference between the measured relative frequency and the true probability. For example, we can determine the approximate probability distribution of a quantum particle $P(x)$ by repeating a position test on many identically prepared particles. More particles are involved, more accurate will be the probability estimate. Each particle is detected with a random position whose statistical distribution is given by $P(x)$.

In cognitive psychology, it is necessary to carefully define the concept of test. Given a preparation with an incomplete information, the subjects' answers in a cognitive experiment are clearly stochastic. But this does not mean that subjects give a random answer just like quantum particles. Usually, cognitive tests ask subjects to give the answer they think is correct, not to give the first answer that comes into mind. Moreover, in other experiments subjects are asked to judge the probability of an unknown event. This means that cognitive tests are less direct than tests in quantum physics, and they are usually designed to exctract more information from subjects, even if it is the result of a cognitive elaboration.

A second important differece is that in cognitive experiments the subject's answer can be considered not only as an information about the cognitive state before the test, but also a new information. In fact, the answer itself is an event, and the subjects, when they have enough cognitive resources, are aware of their answer and they consider it in successive judgements and decisions. This is different from tests in quantum phsyics, where the system dimension does not change after the experiment. 
%
%

%%%%%%%%%
\section{Quantum knowledge postulates for cognition}
I present here a new set of postulates for the new quantum cognition model. They differ from those defined in \cite{busemeyer2011quantum} in many points, even if the mathemathics is the same. This new model will be named \textit{quantum knowledge model of cognition}(QKT), while the other \textit{quantum decision model} , underlining the fact that the two models give predictions about different things: the former about the knowledge of uncertain events (mainly judgements and indirectly choices) while the latter only about decisions.
\\
\\
\textit{Postulate 1: subjects' cognitive representation is given by a state vector in a multidimensional space (technically, an N-dimensional Hilbert space). The architecture of the managed information determines the type of representation, using the minimum of cognitive resources.}

The first postulate introduces the concept of cognitive representation involved in a cognitive task. When asked to perform a task, subjects use a cognitive representation which allows to consider and elaborate the known information. 
The postulate introduces two important new concepts, the \textit{ information architecture (or structure)} and the \textit{cognitive resourses}.

Let us first consider the information architecture. In the quantum formalism, a composite system is given by the tensor product of the Hilbert spaces of the subsystems $H_1 \otimes H_2 \otimes ...$.  Each Hilbert space is used for representing a particular type of information. The information structure that subjects have to manage strongly determines the structure of the Hilbert space involved in the cognitive representation. However, in general,  subjects can interpret their representations in different ways, leading to simpler or richer structures. Let us consider some examples, starting with the basic configuration of two dichotomic events $A$ and $B$. For instance,  $A$ \textit{Linda is feminist} and $B$ \textit{Linda is bankteller}, like in the classic conjunction fallacy experiments. The actual quantum-like models allow building two different configurations. The simplest is a bi-dimensional space $H$ (\textit{non-degenerate model}) \cite{franco2009conjunction}, where the two dimensions correspond to the event \textit{Linda is feminist} and to its negation. This model is based on a single qubit. On the contrary, subjects could consider the description of Linda and build a more detailed representation in the working memory. In this case, the dimensionality of the Hilbert space is higher than 2 and the model is \textit{degenerate} \cite{busemeyer2011quantum}. This second model is supposed to be used whenever subjects have more time to think. In general, however, nothing prevents subjects from using two different Hilbert spaces for such events (two qubits or two different registers relevant for example to Linda's interests and jobs respectively), leading to a composite representation.

What determines the subjects' strategy? Of course the time to answer, but also the information architecture. In the Linda experiment, the two events are relevant to the same person and they are somehow correlated. In other cases, instead, it is more natural to represent the two events in different spaces. For example, unrelated events (like balls drawn from two different urns in \cite{wedell2008testing}) could be better represented with two different Hilbert spaces. In fact, an event relevant to the first urn has no correlation with an event relevant to the second urn. Composite representations could be necessary also when the information is structurally different and can be enriched in time. For example subjects could have to manage decisions and judgements about an uncertain event, like in post-decisional dissonance experiments \cite{knox1968postdecision}, where bettors are asked to decide which horse will win and after to judge the probability that such horse will win. In this case, the basic representation (the horse wins or looses) is enriched with the decision representation (the previous bet on the horse). The introduction of additional quantum registers - a quite standard strategy in quantum computation - thus becomes important also in the context of quantum cognition. In \cite{pothos2009quantumBAE} we can find an example of a composite system to manage two different kinds of choices, but this is in the context of quantum decision theory and it doesn't seem part of a general strategy.

The second point, introduced by postulate 1,  states that subjects tend to use the minimum of cognitive resources in order to define their representations. This is consistent with the studies  integrating psychology and formal information theory in terms of storage capacity. Supposing that one bit of information discriminates between two alternatives, experimental results support the "$7 \pm 2$ bits" theory of storage \cite{miller1956magical}. In other words, subjects have to face with a limited storage capacity for their working memory, unless considering chuncking strategies (which will be treated in a separate paper). It is clear however, that if subjects have to produce judgements in very short time, the avaliable storage capacity will be very low. On the contrary, more time to think will lead to the availability of more resources and thus more complex information structure.

We can now recall the concept of pure state.  If a quantum system is prepared in such a way that it certainly yields a predictable outcome in a specified maximal test, the various outcomes of any other test have definite probabilities. In particular, these probabilities do not depend on the details of the procedure used for preparing the quantum system, so that it yields a specific outcome in the given maximal test. A system prepared in such a way is said to be in a pure state.

The simplest method for producing quantum systems in a given pure state is to subject them to a complete test, and to discard all the systems that did not yield the desired outcome. For example, perfect absorbers may be inserted in the path of the outgoing beams that we do not want. When this has been done, all the past history of the selected quantum systems becomes irrelevant. The fact that a quantum system produces a definite outcome, if subjected to a specific maximal test, completely identifies the state of that system, and this is the most complete description that can be given of it.

Identical preparations for different subjects allow using the same vector state $|\psi\rangle$ as described by the first postulate. Such state is also called \textit{pure state}. Different preparations require to use \textit{mixed states}, which are statistical mixtures of pure states with weights represented by probabilities relevant to the preparations themselves. They are described mathematically by the density matrix $\rho$.

It is important to stress that receiving the same information does not entail the same cognitive representation for different subjects. In fact, they could have different mental models, enabling different ways to elaborate and encode such information.
\\
\\

\textit{Postulate 2. The certain knowledge about a verifiable event is represented by a subspace of the vector space and thus by a projector.} 

The formulation of the second postulate is different from that defined in \cite{busemeyer2011quantum} because it underlines that the subspace is not made in correspondence with an event, but with the certain knowledge of an event.
The actual quantum decision model contains an ambiguity about the concept of \textit{event} that leads to some  problems. Let us analyze this ambiguity in detail. Given a question like \textit{Is Linda feminist?}, we have to define two distinct events: \textit{Linda is feminist} and \textit{The subject says that Linda is feminist}. In other words, subject's answer "yes" to a question represents another event which is distinct from the original event described in the question.
In the original postulate, there isn't such distinction and the simple act of answering to a question alters the original state.

This ambiguity is made explicity in the post-decisional cognitive dissonance experiments, where the decision about an event has to be considered as another distinct event, which in a second time may alter the judgements about the original event.

I conclude the analysis of the second postulate by noting that the use of projectors, even in a degenerate model, could be generalized in order to produce a more realistic description of reality. In most cases an event cannot be defined in terms of mutually exclusive features, which are required by the actual model. As shown in information architecture, exact organization schemas are less used in real life. Most of organization schemas, for example in the world wide web context, are ambiguous \cite{rosenfeld2002information}. For example, looking for a book may require searching by author or title, but if we don’t know them we need to search by topic, and topics represent an ambiguous organization. Thus it could be necessary to generalize the concept of projectors and use POVM (positive operatov value measurement) operators, as attempted in preliminary articles \cite{POVMkitto2015, POVMkhrennikov2014, POVMmiyaderaquantum}, in order to take into account experimental situations where questions are onn mutually exclusive. A new article about this topic will be submitted \cite{Franco2016POVM}.
\\
\\

\textit{Postulate 3: the judged probability that an event is true equals the squared length of the projection of the state vector onto the subspace representing the certain knowledge of the event. }

Again, the postulate removes the concept of anwser to a question, working only with the knowledge of the event and its degree of knowledge. It is important to underline two important points. The new model is not able to make predictions about the probability that subjects give a particular answer. It works mainly with judged probabilities. Decision probabilities in this model are secondary objects, deriving from the knowledge of uncertain facts. The second consideration is that extracting the judged probability can be considered at a first stage a read-only operation over the cognitive state. However, also the produced judgement becomes part of the cognitive system, because it corresponds to the event \textit{the subject says thay the probability relevant to the event is X}.
\\
\\

\textit{Postulate 4: the updated cognitive state after a subject knows with certainty that an event is true equals the normalized projection on the subspace representing that event}. 

From a formal point of view, the updated state is defined according to the L\"{u}der's rule, like presented in \cite{busemeyer2011quantum}. Thus the consitional state $|\psi_A\rangle$ under the action of projector $\textbf{P}_A$ on $|\psi\rangle$ is $\textbf{P}_A|\psi\rangle / || \textbf{P}_A|\psi\rangle  ||$. It is important to distinguish between the knowledge of a fact and expressing a personal opinion. The new postulate introduces a different way to consider measurement. While in fact a physical system can be measured in order to obtain a value of a physical quantity, a direct measurement of a cognitive system is more difficult. The cognitive state relevant to the description of a possible event changes directly only when new information is provided to increase certainty. Such information can arrive from outside (for example, someone tells us that event is true, or we see directly the event) or from inside (for example, if I make a choice, in the mental space of my choices there is a certain fact).

As we will see later in this article, the post-decision cognitive dissonance experiments can be correctly interpreted with the modified postulate 4 by considering that the choice to bet on an horse does not produce a projection over the initial cognitive state relevant to the event \textit{the horse will win the race}. This projection will happen only when the horse really wins.
\\
\\
\textit{Postulate 5: if no new information is provided, the subject's cognitive state evolves in time through the application of a unitary operation}. 

Thus the free evolution of cognitive states (that is the evolution without answers, judgements, new information) is provided by unitary operators. These are reversible, differently from the addition of new information. The form of the unitary operator may depend on many factors: in section \ref{qgates} such evolution is studied in terms of quantum gates.
\section{The opinion poll problem}
I now show that the new model is able to take into account puzzling problems arising from the previous quantum-like model. In this section I consider the opinion poll problem \cite{khrennikov2014quantum}, where two important properties (the response replicability and the order effect) are shown to be incompatible.

The \textit{response replicability property} states that, for a large class of questions, a response to a given question is expected to be repeated if the question is posed again irrespective of whether another question is asked and answered in between. The \textit{order effect} reveals that the probabilities relevant to subjects's consecutive answers to two different questions change depending on the order of the questions. The main problem in the previous quantum-like model is that each answer produces a collapse on the original cognitive state, which is defined on a unique Hilbert space. Thus the second answers completly removes the information relevant to the first one. The only way to obtain response replicability is to use commuting operators relevant to the two questions, which in turn removes the order effect.

On the contrary, in the quantum knowledge model the decision can produce only indirect effects on the original cognitive state. Each answer leads to the addition of a new Hilbert space describing this new inofrmation. Let us consider the example of the cited article, where  the question $A$ is \textit{Is Bill Clinton honest and trustworthy?} and $B$ \textit{Is Al Gore honest and trustworthy?}. Such questions are defined as non-commuting projectors acting on the same Hilbert space. On the contrary, in the quantum-knowledge model these operators are relevant to the concrete knowledge that Clinton/Al Gore is honest and trustworthy and they act on different Hilbert spaces. Thus the initial cognitive state is, in the simplest case, a pure state given by the tensor product of the states relevant to the uncertain knowledge about Clinton and Al Gore respectively:
\begin{equation}
|\psi\rangle=|\psi_{Clinton}\rangle |\psi_{Al Gore}\rangle
\end{equation}
After the answer to Clinton's question, there isn't new knowledge about Clinton. Only a new information is added (the subject's anwser), thus leading to a new Hilbert space. Moreover, a unitary operator is applied which acts on $|\psi\rangle$ in a conditional way (depending on the anwser about Clinton). This unitary operation is the expression of a cognitive-dissonance reducing strategy, and it consists in a rotation of the original belief state towards the new state.
\begin{equation}
|\psi'\rangle=|a(i)_{Clinton}\rangle U_i |\psi_{Clinton}\rangle |\psi_{Al Gore}\rangle
\end{equation}
where $a(i)_{Clinton}$ is the answer to Clinton's question, and $i=0,1$ (false/true). If the first question would be the one relevant to Al Gore, we should have
\begin{equation}
|\psi''\rangle=|b(j)_{Al Gore}\rangle U'_j |\psi_{Clinton}\rangle |\psi_{Al Gore}\rangle
\end{equation}
It is clear from this formalism that the two unitary evolutions $U$ and $U'$ could be different and thus lead to the question order effect. Cognitive dissonance experiments suggest that such evolutions will confirm the judgements relevant the first answer and adapt the uncertainty relevant to the second question accordingly. However, the  replicability property is preserved, because the information relevant to the first anwser is encoded into a different quantum register (relevant to the first Hilbert space). 
\section{Postdecision cognitive dissonance}
In this section, I present another new case where the previous quantum-like model evidences some issues, the postdecision cognitive dissonance experiments. 

Cognitive dissonance theory, first developed by Leon Festigner \cite{festinger1962theory}, dominated social psychology research form the 1950s until the 1970s. The theory revoluzioned thinking about psychological prcesses, particularly regarding how rewards influence and attitude behavior, and how behavior and motivation influence perception and cognition. According to the theory, when an individual holds two or more elements of knowledge that are relevant to each other but inconsistent with one another, a state of discomfort is created, called \textit{dissonance}. Persons are motivated by such state and they may engage in \textit{psychological work} to reduce the inconsistency, for example by adding consonant cognitions, increase the importance of consonant cognitions or decreasing the importance of dissonant ones.
In the 1990s, research on the theory was revived and it has since been gaining in interest \cite{wicklund2013perspectives, harmon2012cognitive, cooper2011cognitive, cooper2007cognitive, greenwald1978twenty}.

The name itself of cognitive \textit{dissonance} looks like very familiar in the context of quantum cognition, suggesting something like an ondulatory behavior. In fact, it has been already considered in \cite{mogiliansky2009type} where we can find a first attempt to design a quantum-like model for cognitive dissonance. However, this model only considers dissonance between decisions, using a simple model where the decisions correspond to non-commuting observables on the same Hilbert space. Such a simple model of course leads to the same problems already described for the opinion poll problem, violating the replicability principle. Moreover, the cognitive dissonance has also been considered in the context of quantum cognition to explain the violation of the sure thing principle of decision theory \cite{pothos2009quantum}, even if the generality of the mechanism seems not to be fully evidenced.

Postdecision cognitive dissonance is a particular case of cognitive dissonance, where dissonance is produced between  decisions and a subsequent test. This final test can be a preference choice \cite{sharot2010decisions} or a judgement \cite{knox1968postdecision}. In particular, results in \cite{knox1968postdecision} show that bettors at a horse racetrack are more confident in their chosen horse just after placing the bet because they cannot change it (the bettors felt \textit{post-decision dissonance}). In other words, the commitment to a decision activates dissonance-reducing processes which modify previous judgements. Various experiments confirm this result: for example, voting for a candidate leads to a more favorable opinion of the candidate in the future \cite{beasley2001cognitive, mullainathan2009sticking}.
This effect seems to be very general and powerful. For example Deutsch and Gerard \cite{deutsch1955study} found a similar result in a social validation context: when people have to make a judgement or a choice, the simple act of writing the decision becomes a commitment and influences judgements.

Let us now briefly analyze these experimental results in more detail. Subjects' decisions, if not explicitly performed, are reversible: subjects can change idea. On the contrary, subjects making an explicit choice make a conscious and irreversible act. For example subjects could tell to someone their choice or write it on a paper. In any case, their choice or judgement is now a known and certain fact, it cannot be modified anymore (of course they can change, but they have to declare the first choice as wrong). This is different from simply thinking to a choice or a judgement.  This analysis is consistent with the postulates previously presented. 
In particular, I underline two points. 

First of all, subjects' decision (the chosen horse) \textit{become a new part of the cognitive system}. In the quantum-like formalism, we can say that the act of making an explicit choice creates into the cognitive system a new element in a new Hilbert space. Thus we have started with a single cognitive system, relevant to the judgement of a possible event (e.g. the horse will win the race) and we have now an additional cognitive system relevant to the taken choice. These two subsystems form together the subject's cognitive system and they need to be considered together. The study of only one of them makes impossible in general giving a correct description of subject's behaviour. This new point has been described in the first postulate. 
In the horse-race example, the two subsystems  are in different states. While the first describes uncertainty about the race, the new subsystem represents a certain knowlegde about the chosen horse. This last state cannot be modified anymore, it is certain and non-reversible. Its influence on subsequent judgements may depend from many factors, like for example the cognitive dissonance or the social validation. In the last part of the article, we will see how the quantum formalism allows describing such influence.

The second point to underline is that \textit{only new knowledge produces collapse}. The postdecision dissonance experiment is particularly interesting because it allows to test the validity of postulate 4 in  the quantum-like model. Let us consider for semplicity an event $A$, which can be true or false. For example, $A$ can be \textit{that horse will win the race}. According to postulates in \cite{busemeyer2011quantum}, if $A$ is concluded to be true, the initial state describing subject's judgement is projected by using a suitably defined projection operation. This postulate is analogue to the postulate of quantum physics also called \textit{collapse postulate}. According to it, a measurement causes a sudden and non-reversible change of the physical state into a new state describing the certain knowledge of measured physical quantity. On the contrary, when the system is not measured, its evolution in time is driven by suitable unitary operators and it is  always reversible. According to the previous quantum decision model \cite{busemeyer2011quantum}, subject's decision to bet on a horse would lead the subject to change its mental state. In the post-bet condition, the subject should simply judge that the winning probability for the chosen horse is 100\%, which is the state consistent with the fact. This is of course a non-realistic description. Where is the problem? The only case where we can think that the collapse happens is when the subject \textit{knows} that an event is true. For example, after the end of the race, the subject knows which horse has won and its cognitive state represents such certain event. On the contrary, the previous version of the quantum cognition model assumed that the collapse just happens when the subject makes a choice, even if he does not really know what is the correct answer.
Just like described for the question order effect, the cognitive system becomes bipartite: one subsystem is relevant to the knowledge of the bet decision, the other is relevant to the knowledge about the horse (the probability to win). After the bet, the system is subjected to a conditional unitary operation which modifies the second subsystem. It consists essentially in a rotation of that state towards a state consistent with the taken decision. This is a dissonance-reducing strategy.
\section{The belief-action entanglement model}
It is important to distinguish this approach from the \textit{belief-action-entanglement} model (BAE) presented in \cite{pothos2009quantumBAE}. In this paper the Hilbert space used to represent the experiment is composite. The first subsystem is relevant to the categorization, while the other about the action: in both cases, they can be considered as decisions, and thus the model can be more generically considered as a decision-decision entanglement model.

First of all we note that further research should focus on the link between judgement and decision. However, as we have seen in the previous sections, a quantum model based on decisions can lead to some problems when we add judements to the experiment. Let us suppose that after the categorization-decision test in \cite{pothos2009quantumBAE} subjects are asked to produce a probability judgement about the fact that the face is relevant to a good/bad person. According to the actual model, the previous decision should impose a probability judgement of 1 (certainty about the good/bad person). Of course this is not consistent with real facts. At least subjects would adapt their judgements to the previous chioces, but they wouldn't show a collapse effect.

We could also imagine another variant of the experiment, where subjects are again asked to perform the initial categorization task. According to the BAE model,  \textit{the potential to categorize the face as a good guy is determined by an amplitude $\psi(G)$, and if this response is obtained, then it would produce a transition to state G}. Thus the new categorization task will simply produce again the same result, beacuse the state is collpased on the answer $G$.
On the contrary, subjects could make a different categorization from the previous one,  because their cognitive state about the good/bad categorization is uncertain. This could be true expecially for ambiguous faces.
\section{Grand repiprocity equations}
I now present an important new prediction of the quantum cognition model that seems to change the actual conclusions about quantum-like models.

Non-degenerste models show a very simple property, expressed by the \textit{grand-reciprocity equations}, also known as GR equations. Given two non-commuting observables $A,B$ acting on a single-qubit Hilbert space, we have that 
\begin{eqnarray}
p(A_0|B_0)=p(B_0|A_0)=p(A_1|B_1)=p(B_1|A_1)\\
p(A_0|B_1)=p(B_1|A_0)=p(A_1|B_0)=p(B_0|A_1)
\end{eqnarray}
However, an experimental test has been made in \cite{boyer2016testing} by considering answers to two dichotomic questions. The test evidenced violations of the GR equations. 

According to quantum decision model, agent's belief state just after the answer is the normalized projection of the belief state onto the vector corresponding to the answer. The test thus computes the probability that many different subjects give the same answer (for example TRUE) to a question.  We can say that such quantum-like model works with answer probabilities: it is in fact a quantum decision model.

The model cannot say anything in general about mean judged probabilities without additional hypotheses. Of course one can assume that judgements follow direct answers, but in general this does not allow to derive directly a functional link between them. For example, subjects could give a mean 70\% judged probability about Clinton honesty with low variance, leading to a 100\% positive direct answer.

I thus conclude that the experimental test is only relevant to the quantum decision model. Grand reciprocity equations, when tested  with such probabilities, are violated: answer probabilities are not fully symmetric.

The new model I propose instead works with judged probabilities. As also recognized by one of the authors of \cite{boyer2016testing}, at the moment there aren't experimental test relevant to the symmetry of judged probabilities. Thus we can say that non-degenerste models are at ther moment still valid in the context of quantum cognition models. Further research could evidence that such models give good results only in spceific experimental conditions, like fast judgements and correlated events. This seems to find a first confirm from the inverse fallacy experiments \cite{franco2007inverse}, where subjects manifest symmetry in judging conditional probabilities. Moreover, also similarity and typicality judgements generally manifest symmetry. Quite interestinly, the symmetry violations of similarity have found an explanation in a quantum-like framework by considering a non-degenerate model \cite{pothos2011quantum}. This suggests that symmetry is relevant to situations where the judgement is performed with low mental resources, while asymmetry to situations requiring a deeper analysis.

\section{A quantum computational approach to quantum cognition}\label{qgates}
The use of concepts and operations taken from the quantum computation field may allow the quantum cognition model to produce new results and predictions. In fact, the use of quantum computational gates can help to find basic patterns in the unitary evolution of the system.
Previous attempts to use quantum algorithms in the context of quantum cognition (see for example \cite{franco2008grover,franco2009availability, pleskac2015computation}) show - only in part - the potentialities of this approach. 
In particular in \cite{pleskac2015computation} we can see the use of quantum computational gates, while in  \cite{franco2008grover,franco2009availability} we can find some attempts to use algorithms derived from Grover's algorithm to model some memory features.

Following  \cite{franco2007quantum}, we first consider a single \textit{qubit} model, where the cognitive task is represented by a dichotomous question A: this question can be formulated in such a way that it admits only two mutually exclusive answers, “yes/true/1” and “no/false/0”. 
In the revisited quantum cognition model, vector $|1\rangle$ represents a situation where question about A has a \textit{certain} true answer.In a similar way the vector $|0\rangle$ is defined. We underline the fact that even if a subject chooses the 1 answer, the cognitive state is $|1\rangle$ only if there is a certain knowledge that the answer to $A$ is true.

The inner product between a generic state $|s\rangle$ and $|1\rangle$ is the amplitude $\langle 1|s\rangle$, and its squared magnitude, $|\langle 1|s\rangle|^{2}$, corresponds to the probability that the answer to A is ‘yes’ with certainty. In other words, it represents the partial subject's knowledge.  Thus, according to the quantum formalism, the opinion state $|s\rangle$  can be in a linear superposition of vectors $|1\rangle$ and  $|0\rangle$, where the superposition coefficients are the amplitudes $\langle 0|s\rangle$ and $\langle 1|s\rangle$. This superposition represents an indefinite state.
\\
In the two-qubit case, states are vectors in a space obtained by the tenson product of the consituent Hilbert spaces. For example, the vector $|00\rangle$ means that we have the tensor product of two single qubit vectors in 0 state.

A quantum logic gate is a unitary operator which reproduces in the quantum context the behaviour of a logic gate. As shown in Deutsch \cite{deutsch1989quantum}, we can define a \textit{universal set of gates}, in the sense that every quantum algorithm can be efficiently simulated to arbitrary fixed accuracy by a circuit composed by such gates.

\subsection{Single qubit gates}
We first consider the universal set of gates acting on a single qubit, which is formed by the Hadamard gate $H$ defined as follows
\begin{equation}
H|0\rangle=\frac{1}{\sqrt{2}} (|0\rangle + |1\rangle),  
H|1\rangle=\frac{1}{\sqrt{2}} (|0\rangle - |1\rangle)	
\end{equation}
and by the phase shift gate $R(\phi)$
\begin{equation}
R(\phi)|0\rangle=|0\rangle,  
R(\phi)|1\rangle=e^{i\phi}|1\rangle	
\end{equation}
where the phase of the vector $|1\rangle$ is changed, and $i$ is the complex number. 
\\
An interesting operator, obtained by the combination of such gates, is the \textit{single-qubit interference operator}, definbed as $I(\phi)=HR(\phi) H$, which simulates the Mach-Zehnder interferometer:
\begin{equation}\label{mach}
I(\phi)=HR(\phi) H|0\rangle=cos(\phi/2)|0\rangle + i  
sin(\phi/2)|1\rangle 
\end{equation}
The $|0\rangle$ vector is conventionally taken as a reference starting state. Every quantum algorithm normally starts from it. The single-qubit interference operators can by simply combined in the following way: 
\begin{equation}\label{sum_interference}
I(\phi)I(\theta)=I(\phi+\theta)
\end{equation}
This operator has a very interesting interpretation: the initial state $|0\rangle$ is split into an equal superposition with the first Hadamard operator. Then, a phase shift is introduced so that, when recombining the states with the final Hadamard transform, interference terms appear. Of course a null phase shift $R(0)$ reduces the operator to $HH$, which is the identity operator.

Let us now try to use these concepts in the quantum cognition context. The Hadamard gate is a representation change between two statistically unrelated events $A$ and $B$ (that is  $P(A|B)=1/2$ and $P(\neg A|B)=1/2$). It is not important at this stage to identify the meaning of the new representation, we can simply say that is an unconscious change of viewing things. Two consecutive Hadamard gates ($HH$) describe a change followed by a return to the initial representation, and so it is equivalent to the identity.
The operator $HR(\phi) H$ is very interesting also in the quantum cognition context. We can interpret it as  \textit{the basic thinking activity}, which leads to  a change in beliefs. It is a representation change followed by a phase shift, with a final return to the inital representation.
\\
What is the meaning of the phase shift? It is evident from the single-qubit interference operator that it has a very important role, since the magnitude of the phase shift determines the intensity of the interference. First of all we note that it has not an equivalent in the classic world. Equation \ref{mach} evidences that we can have positive or negative interference, depending on the value of the phase. 

Let us now try to use these concepts in the postdecision experiment. Let belief state $|s\rangle=I(\phi_0)|0\rangle$ describes a situation where the subject makes his own judgement about the horse. If $\phi_0=\pi/4$, the estimated probabilities that the horse will win (or not) are the same.
Thus the application of the single-qubit interference operator $I(\phi)$ to $|s\rangle$ will change the subject's judgement, making him more or less confident about the fact that the horse will win. What determines this? Simply the relative phase between the inital phase and the phase shift. The final resulting phase will be $\phi_0 + \phi$. If $\phi$ is positive and less than $\pi/4$, the probability relevant to $|1\rangle$ increases. If $\phi$ is negative but higher than $-\pi/4$, the probability relevant to $|1\rangle$ decreases. In the next subsection we analyze coditioning factors that can control such relative phase.
\subsection{Two qubit gates}
One simple set of two-qubit universal quantum gates is the Hadamard gate (H), the phase shift gate , and the controlled NOT gate. The controlled NOT gate, or C-NOT is defined as
\begin{eqnarray}
CNOT|00\rangle=|00\rangle, 
CNOT|01\rangle=|01\rangle,\nonumber\\
CNOT|10\rangle=|11\rangle,
CNOT|11\rangle=|10\rangle
\end{eqnarray}
Tthe combination of such gates allows obtaining a particular kind of two-qubit gates, the conditional gates. For example, the \textit{controlled phase shift} $CR(\phi)$ is defined as
\begin{eqnarray}
CR(\phi)|00\rangle=|00\rangle, 
CR(\phi)|01\rangle=|01\rangle,\nonumber\\
CR(\phi)|10\rangle=|10\rangle,
CR(\phi)|11\rangle=e^{i\phi}|11\rangle
\end{eqnarray}
which evidences that the phase shift is applied to the second qubit only when the first qubit is in state 1.
We can also write a \textit{conditional single-qubit interference operator} (or simply conditional interference operator), which acts as \ref{mach} only if the first qubit is $|1\rangle$.

The conditional interference operator is interesting in the quantum cognition context. In the postdecisional cognitive dissonance experiment, the second qubit (conventionally) describes the judgement about the horse race, while the first qubit is relevant to the explicit decision to bet. Before betting, the first qubit is an indefinite state. But after betting (or publicly choosing an horse), this qubit is in a definite state (for convention, $|1\rangle$). Now the conditional interference operator acts modifying the judgements about the horse. The phase shift derives from the cognitive dissonance. It is easy to understand that the choice ($|0\rangle$ or $|1\rangle$) will activate two different conditional phase shifts. We can define a general conditional phase shift $CR(\phi_0,\phi_1)$ as
\begin{eqnarray}
CR(\phi_0,\phi_1)|00\rangle=|00\rangle, 
CR(\phi_0,\phi_1)|01\rangle=e^{i\phi_0}|01\rangle,\nonumber\\
CR(\phi_0,\phi_1)|10\rangle=|10\rangle,
CR(\phi_0,\phi_1)|11\rangle=e^{i\phi_1}|11\rangle
\end{eqnarray}	
Thus the state before betting is
\begin{equation}
|s_0\rangle=\frac{|0\rangle+|1\rangle }{\sqrt{2}}R(\theta)|0\rangle
\end{equation}
where $R(\theta)|0\rangle$ is the conventional way to describe the initial undefined state. The angle $\theta$ can be $\pi/4$ in case of complete uncertainty. The state after the betting is $CR(\phi_0,\phi_1)|s_0\rangle$, which results
\begin{equation}
\frac{|0\rangle R(\theta+\phi_0)|0\rangle + |1\rangle R(\theta+\phi_1)|0\rangle}{\sqrt{2}}
\end{equation}
It is easy to see that when $\theta=\phi_1=\pi/4$ and $\phi_0=-\pi/4$ we have a maximally entangled state. This is of course and extreme case, where the act of betting strongly influences judgements. A positive decision (value 1 in the first register) leads to a $\pi/4$ additional shift. On the contrary, a negative decision (0 value in the first register, that is \textit{the horse won't win}) leads to a $-\pi/4$ decision.

We can finally try to provide an interpretation for the phase shift operation in terms of cognitive dissonance: subjects, when managing different registers relevant to the same event, perform unitary operations trying to align such registers. This is spontaneusly driven by the  mental stress or discomfort experienced by a subject, which in the quantum cognition model corresponds to the phase shift contained in the conditional operator $CR(\phi_0,\phi_1)$.
If the perceived cognitive dissonance (the feeling of stress) is low, this means that $\phi_0=\phi_1=0$, and the conditional operator is the identity operator: the decision encoded in the first register doesn't entangle with the second register. A strong dissonance, on the contrary, corresponds to a strong phase shift ($\phi_0=-\pi/4,\phi_1=\pi/4$).
This interpretation is also consistent with the description of phase provided in the Grovers' description of human memory \cite{franco2008grover}, where the efficiency of the algorithm is shown to be determined by the value of phase shift. Just like the search algorithm works better with a particual value of the phase shift, similarly the entity of the cognitive dissonance effect can be put in correspondence with the value of the phase shift involved in the quantum operators relevant to the evolution.
%
%%%%%%%%%%%%%%%%%%%%%%%%%%%%%%%%%%%%%%%%%%%%%%%%%%%%%%%%%%%%%%%%%%%%%%%%%%%%%%%%%%%%%%%%%%%%%%%%%%%
\section{Conclusions}
In this article I present a new approach to the quantum cognition, where a more coherent use of the concept of state collapse is provided. In fact, given an uncertain event, it is evident from postdecision cognitive dissonance experiments that choosing the most probable outcome does not entail a collapse of the cognitive system. The only way the collapse may happen is when new information is given and there is certainy about the event. Thus postulate 4 of the quantum cognition model is modificed and consequences of it are derived.

In particular, it results that even the simplest cognitive operations lead to the creation of composite systems in the quantum cognition model. To treat them, it is usefult to use concepts derived from the quantum computation, like for example conditional operations and entangling gates.

Finally, I underline that this new approach needs further work and test with different cognitive experiments. The conditional operations could be very useful in studying social validation experiments, where information provided by other people influence the single subject. An approach using concepts taken from spin chains models (Isin model for example) could lead to interesting results, treating such effects  like phase transitions. 
\section{Appendix}
A quantum system is said to be \textit{composite} when it is
composed by different subsystems (for example, different
particles). In particular, in the case of two subsystems the state
is said \textit{bipartite}. The simplest example of a bipartite
state is given by a two-qubit system. A two-qubit pure state is a vector in a space obtained by the tenson product of the consituent Hilbert spaces.
For example, the vector $|00\rangle$ means that we have the tensor product of two single qubit vectors in 0 state. 

A bipartite state is said to be \textit{separable} or \textit{factorable} if it can be written in the form
\begin{equation}\label{factorable}
\rho=\sum_i p_i \rho'_{i}\otimes \rho''_{i}
\end{equation}
where $\rho'_{i}$ and $\rho''_{i}$ are density matrices relevant to the first and the second subsystem, respectively. In the special case when $\rho=\rho' \otimes \rho''$ the density matrix is named {\it product state} and represents a particular case of separable state. In general, separable states can exhibit classical correlations.

Entanglement is one of the most intriguing features of quantum systems. A state is defined as \textit{entangled} if it cannot be written as in formula \ref{factorable}.
Entangled states exhibit correlations between the subsystems that are stronger than the analogous correlations in the classic case.  
The simplest entagled pure states are the Bell states, which are also maximally entangled states:
\begin{eqnarray}
|\Phi ^{+}\rangle ={\frac  {1}{{\sqrt  {2}}}}(|00\rangle +|11\rangle )\\
|\Phi ^{-}\rangle ={\frac {1}{\sqrt {2}}}(|00\rangle -|11\rangle ) \\
|\Psi ^{+}\rangle ={\frac {1}{\sqrt {2}}}(|01\rangle +|10\rangle )\\
|\Psi ^{-}\rangle ={\frac  {1}{{\sqrt  {2}}}}(|01\rangle -|10\rangle ).
\end{eqnarray}
\footnotesize
%
%%%%%%%%%%%%%%%%%%%%%%%%%%%%%%%%%%%%%%%%%%%%%%%%%%%%%%%%%%%%%%%%%%%%%%%%%%%%%%%%
%
\bibliography{../mybib}{}
\bibliographystyle{unsrt}
\end{document}